# E-commerce between a large firm and a SME supplier: a screening model

Alderete María Verónica

**Abstract**—This paper derives a model of screening contracts in the presence of positive network effects when building an electronic commerce network (e-commerce) between a large firm and a small and medium sized enterprise (SME) supplier based on Compte (2008). Compte (2008) main insight is that when several potential candidates compete for the task, the principal will in general improve the performance of his firm by inducing the member candidates to assess their competence before signing the contract (through an appropriate choice of contracts). The large firm (principal) must choose between different SME suppliers (agents) to build a business to business e-commerce network. In the presence of positive network externalities, we show that social surplus increases.

**Index Terms**—Economics, Electronic Commerce, Model theory, Software acquisition.

——————————— ◆ ———————————

## 1 INTRODUCTION

There are many types of production networks and some experiences show that e-commerce enforces network activities. The higher the relative benefits offered to customers and partners, the higher their incentives to stick with or join the network established by the e-business. Following [1] the study considers that firms value being part of a large network, i.e. using an ICT (information and communication technology) business tool that many other firms also use. This is the direct network effect. Firms also value a hardware technology for which there is a wide variety of software available, and more software's firms associate with a hardware technology if more firms use it. This is the indirect network effect. For example, windows operative system is widely used among firm's computers, because a lot of management software uses windows, instead of other operative system, as MAC.

In the presence of network effects [2], the utility from adoption increases in the number of other adopters that purchase or introduce the innovation. Interdependence is a powerful source of positive feedbacks among adopters who take into account both the current network size and its perspective for future growth. Buyer supplier electronic commerce means finding a partner with which the firm can establish a bilateral relationship and having the partner undertake relationship-specific investments so that it becomes able to respond electronically to the firm's particular needs.

The increasing return properties inherent to network effects then magnify the relative benefits offered, thus triggering positive feedback dynamics. For example, there are networks with a leader company, usually a large company that establishes the membership requirements. In traditional supply chain mass production oriented, as in the automobile industry, subcontractors are generally "mono clients" and network membership is relatively fluid. In this situation, e-commerce can reduce communication costs and critical times, without altering power relationships inside the network.

This paper develops a simple principal agent model to explain a buyer-supplier e-commerce network building. It analyses the decision of a large firm about contracting an innovative SME supplier, to build and manage an ICT business tool, as electronic commerce. The suppliers are planning to modernize their businesses by incorporating ICT, but they could not do it alone, due to financial constraints and technical assistant support needs. Thus, contracting with the large firm is the only way for a SME supplier to introduce an ICT business tool. Contracting problems between a large firm and her SME suppliers limit the types of contracts that can be written between them. The large firm search for a particular supplier partner to build electronic commerce. The study finds that network effects related to technology compatibility as well as informational spillovers increase the social expected surplus.

## 2 THE MODEL

Following [3] a large firm (principal) must choose among different supplier firms to build a B2B electronic commerce network. The principal could be a large firm that decides to initiate electronic commerce with one of her suppliers. The principal has to choose among her suppliers to introduce some ICT business tool for technology compatibility and provide them the right incentives to perform this task.

Compte and Jehiel(2008) main insight is that when several potential candidates compete for the task, the principal will in general improve the performance of his firm by inducing the member candidates to assess their com-

————————————————

• *María Verónica Alderete works at the National Commission on Scientific and Tecnology Research (CONICET) and the Department of Economics, Universidad Nacional del Sur, (8000) Bahía Blanca, Argentina.*

petence before signing the contract (through an appropriate choice of contracts). The manager will (at least in some cases) be better off when candidates can more easily assess their competence, hence for example when he provides a more accurate description of the job to the agents.

In essence, their result follows from the following observation: when several agents are in competition for the contract, information acquisition accompanied by a proper screening device is socially desirable because it increases the chance that the principal will pick the most appropriate agent. In this study, it is more likely that the manager will pick the most appropriate supplier firm in terms of technological compatibility to engage into the e-commerce network.

In their setup, it is not merely the fact that there are several agents that drives the result. It also requires that private information bears on agents' competence (which is assumed to be agent specific) rather than on common characteristics that would apply equally to all agents.

Because we talk about goods that have network externalities, the utility of the agent will depend also on the network good.

The large firm attaches a value of V(q) to the adoption of q ICT business tools. In this model, firms are supposed to adopt different ICT business tools. Assume that V(0)=0, and let V(1)=v and V(2)=V.

Thus, q= 0, q1, q2.
- q1=1 means the firm adopt a basic ICT business tool.
- q2=2 means the firm adopt a more sophisticated ICT business tool.

This ordinal output assumes that ICT business tools are different in quality; thus, the principal or manager attaches a different value to the acquisition of them.

There are n supplier firms, all of them have planned to incorporate ICT business tools and the chosen one will incorporate these technologies as a task. For each agent i, the disutility of adopting an ICT business tool is equal to $\beta_i q$, where $\beta_i > 0$. $\beta_i$ are the desutilities of technology adoption, each lower $\beta_i$ indicates that the firm candidate i fits the requirements better. It happens that some firms are more internet-oriented than others, such as a firm with qualified employees and entrepreneur skills that foster a better technology adoption.

Because of ICT network externalities, there is an additional benefit for the supplier firm (additional to the monetary payment specified in a contract for non network goods) that comes from the size of the network good $q_i$ that is defined as the total amount of network good in this buyer-supplier relationship, usually referred as network size. Then, $q_i$ is at the same time the total amount of the network good the supplier firm consumes.

Utility is transferable. If $q_i$ ICT tools are introduced and the principal pays t to agent i, the payoff to the principal is V(q)−t and the net benefit to agent i is t−($\beta_i$-γ)q, where γ is a parameter that reflects the preferences of the supplier over the network good. Thus, the net benefit of the agent depends negatively on the disutility of technology adoption and positively on the utility of the network good.

The parameters of the problem are known to all parties, except for the desutilities of production, $\beta_1,\ldots,\beta_n$; and the utilities of the network good $\gamma_1,\ldots,\gamma_n$, which are supposed to be identical among agents. We assume that each $\beta_i$ can take two possible values, $\beta_i \in \{\underline{\beta}, \overline{\beta}\}, 0 < \underline{\beta} < \overline{\beta}$ and we assume that these parameters are drawn from identical and independent distributions. We let p denote the probability that firm i is a low cost agent: p=Pr{$\beta_i = \underline{\beta}$} and (1-p) the probability of being a high cost agent and assume that 0<p<1.

The social surplus is equal to S=V(q$_i$) −( $\beta_i$-$\gamma_i$)q$_i$

Following [3] assume that for a low cost agent, the social surplus is largest when a sophisticated ICT tool is adopted, and that for a high cost agent, the social surplus is largest when a basic ICT tool is adopted, that is:

For $\underline{\beta}$, S = V-2($\underline{\beta} - \gamma$)

If V-2($\underline{\beta} - \gamma$) > v-($\underline{\beta} - \gamma$), then V-v > ($\underline{\beta} - \gamma$)

For $\overline{\beta}$, S = v-($\overline{\beta} - \gamma$)

If v-($\overline{\beta} - \gamma$) > V-2($\overline{\beta} - \gamma$), then ($\overline{\beta} - \gamma$) > V-v

As a result, ($\underline{\beta} - \gamma$) < V-v < ($\overline{\beta} - \gamma$) or $\underline{\beta}$ < V-v+γ < $\overline{\beta}$.

These conditions are equivalent to

$$V - 2(\underline{\beta}-\gamma) > \max\{v-(\underline{\beta}-\gamma), 0\}, and\ v-(\overline{\beta}-\gamma) > \max\{V-2(\overline{\beta}-\gamma), 0\}$$

After signing a contract, the supplier firm will learn at no cost if he is high cost or low cost. In this setting, if the firm observes β immediately after being offered a contract (but yet not signed) it will be at positive cost (c>0) reflecting the difference in cost between acquiring information in the precontractual and postcontractual phases, the weight of the principal assistance over the cost of acquiring firm type information.

The objective of the paper is to expand Compte's results in a network effects setting, to show if these network effects alters the decision of the principal about offering contracts that induce information acquisition.

## 2.1. Results

Following [3], consider the case in which the principal would induce no information acquisition. In this case, candidates are identical ex ante. Thus, it is irrelevant which one is selected (preferences over the network good are identical). The maximum surplus S generated by any match is then obtained when the firm adopts a more sophisticated ICT business tool if low cost and a basic one if high cost. Then,

$$S^* = p(V - 2(\underline{\beta} - \gamma)) + (1-p)(v - (\overline{\beta} - \gamma)) \qquad (1)$$

Compared to the maximum surplus S obtained in a non network goods setting (γ=0), this one is increased by the amount $\gamma(1+p)$. It can be demonstrated that S*=S+ $\gamma(1+p)$. Thus, the higher the probability of being a low cost agent is, the larger the expected surplus will be. Besides, the higher the network good utility is, the larger the expected surplus will be.

Since the agents have the option to refuse any contractual offer, they cannot obtain an expected payoff below 0. Thus, S* is an upper bound on the payoff obtained by the principal when he does not induce information acquisition.

**Proposition I.** *The expected payoff obtained by the principal if he does not induce information acquisition is at most equal to S*.*

According to [4], the payoff obtained by the principal will be strictly smaller than S*, because of the incentives she has to give agents not to acquiring information before signing the contract.

Then, [3] turn to the case in which the principal induces information acquisition. It is assumed that the principal makes a sequence of contractual offers to agent 1, 2, …, n, until one agent accepts.

In this study Proposition 2 generates the same results as in the case of non network goods, the agent prefers acquiring information. The only difference is the payoff to specify in the contract C1, which must add the network externality.

**Proposition II.** *Assume that the principal always offers the contract C, defined as follows: "Introduce a sophisticated ICT tool and receive a transfer equal to $T = 2(\underline{\beta} - \gamma) + c^+ / p$."*
*There exists c'>0 such that if c<c', then (a) any agent who is offered this contract acquires information and accepts the contract if and only if he is a low cost agent, and (b) expected payoff to the principal exceeds S* if n is large enough and γ>0.*

**Proof.** The presence of network effects introduces an additional constraint for the principal's expected payoff to be larger than the expected surplus (n large is not a sufficient condition).

To prove proposition II, observe that a high cost agent will not accept the contract, since $T < 2(\overline{\beta} - \gamma) + c / p$.

If the agent does not acquire information about his type and still signs the contract he gets T-2(Eβ-γ) which is negative for c small enough.[1] Look at the appendix to get this expression.

Following [3] the principal's expected payoff is equal to

$$(1 - (1 - p)^n)(V - T) \qquad (2)$$

Which exceeds S* when n is large enough and c small enough and if the principal internalizes the externality. $(1-(1-p)^n)(V-T)$ converges to $V - 2(\underline{\beta} - \gamma)$ as $n \to \infty, c \to 0$ and because $S^* < V - 2(\underline{\beta} - \gamma)$, since $\gamma >> (v - \overline{\beta}) - (V - 2\underline{\beta})$. The right side of this inequality is negative since

---

[1] For $c < c' = 2 p (1 - p)(\overline{\beta} - \underline{\beta}) - 2 p \gamma$

$V - 2\underline{\beta} > v - \overline{\beta}$.

To prove that $S^* < V - 2(\underline{\beta} - \gamma)$ assume
$V - 2(\underline{\beta} - \gamma) = S^* = p(V - 2(\underline{\beta} - \gamma)) + (1 - p)(v - (\overline{\beta} - \gamma))$
$V - 2(\underline{\beta} - \gamma) = p(V - 2\underline{\beta}) + (1-p)(v - \overline{\beta}) + \gamma(1 + p)$.

Rearranging this expression we get $\gamma = (v - \overline{\beta}) - (V - 2\underline{\beta})$. This can not be true since γ>0, the agent value the network good positively. Following the same reasoning, $\gamma < (v - \overline{\beta}) - (V - 2\underline{\beta})$ is not possible. Then, $\gamma > (v - \overline{\beta}) - (V - 2\underline{\beta})$ and $V - 2(\underline{\beta} - \gamma) > S^* = p(V - 2(\underline{\beta} - \gamma)) + (1-p)(v - (\overline{\beta} - \gamma))$.

### Second Best

Proposition II shows that inducing some information acquisition is good for the principal. To get the optimal contract, the network externalities are introduced, so that the expected payoff obtained by the principal under the optimal contract is function of the information acquisition and the network externality.

Following [3] we define $\underline{c}$ and $\overline{c}^*$, let $\underline{c} = p(1-p)(\overline{\beta} - \underline{\beta})$ and $\overline{c}^* = p(1-p)[V - 2(\underline{\beta} - \gamma) - (v - (\overline{\beta} - \gamma)]$ and observe that $\overline{c}^* = p(1-p)(V - 2(\underline{\beta} - \gamma) - (v - (\overline{\beta} - \gamma))$ can be written as $p(1-p)(V - 2\underline{\beta} - (v - \overline{\beta}) + \gamma)$ which is positive since we have proved that $\gamma > (v - \overline{\beta}) - (V - 2\underline{\beta})$ and $\underline{c} < \overline{c}^*$ as $V - \underline{\beta} > v$.

Define a new contract C2 as follows: "Either introduce a basic ICT tool for a transfer t=Eβ-γ, or a sophisticated ICT tool for a transfer T= t+ $\underline{\beta}(+\varepsilon)$.

**Proposition III.** *Assume that $c \in (\underline{c}, \overline{c}^*)$, n=2 and their definitions of $\underline{c}$ and $\overline{c}^*$. The optimal contract can be implemented by offering C1 to agent 1 and contract C2 to agent 2 in case agent 1 rejects C1. The optimal contract yields to the principal an expected payoff equal to $S^* + \overline{c}^* - c$*

**Proof .**
Step 1: The principal cannot get more than $S^* + \overline{c}^* - c$
The best economic outcome is that when one of the two agents has a low cost he is selected to adopt two improvements and otherwise adopt one. The first event has probability p+(1-p)p, and if the agent acquires information, then the corresponding expected surplus is given by:

$[p + (1 - p) p][V - 2(\underline{\beta} - \gamma)] + (1 - p)^2[v - (\overline{\beta} - \gamma)] =$

$[p+(1-p)p](V-2\underline{\beta})+(1-p)^2(v-\overline{\beta})+\gamma[p+(1-p)p+1] = S^* + \overline{c}^*$ (See appendix). Where S* is the highest surplus obtained with one agent or without information acquisition.

When one agent acquires information the maximum surplus net of information acquisition costs is thus equal to $S*+\bar{c}*$-c, which is the maximun expected payoff the principal can hope to get.

Step 2.

When $c \in (\underline{c}, \bar{c}*)$, the contract C2 yields the principal an expected payoff equal to S*.

**Proof.** By construction we have:

$T - 2(\underline{\beta} - \gamma) > t - (\underline{\beta} - \gamma)$ and $T - 2(\bar{\beta} - \gamma) < t - (\underline{\beta} - \gamma)$. This means $\underline{\beta} - \gamma < T - t < \bar{\beta} - \gamma$.

Hence an agent who would accept this contract without acquiring information would introduce two contents if low cost, and one content if high cost. Then, he would obtain an expected payoff G supplier that satisfies

G supplier = $p(T - 2(\underline{\beta} - \gamma)) + (1-p)(t - (\bar{\beta} - \gamma))$  (3)

This is positive by construction. Besides, for $c > \bar{c}*$, we have $E\beta > \bar{\beta} - c/1 - p$. Hence we have $t > \bar{\beta} - c/1 - p$, so we also have G supplier $> p(T - 2(\underline{\beta} - \gamma)) - c$.

Step 3. When the principal offers contracts as in proposition III he obtains an expected payoff equal to $S*+\bar{c}*$-c.

**Proof.** When the principal offers C1 to agent 1, agent 1 acquires information and rejetcs the contract in the event he is a high cost agent. In the latter case, he offers C2 to agent 2, and obtains an expected payoff equal to S. Overall, his expected payoff is equal to

G principal = $p(V - T) + (1-p)S*$

G principal = $p(V - 2(\underline{\beta} - \gamma)) - c + (1-p)S*$

Since $\bar{c}* = p(V - 2(\underline{\beta} - \gamma) - S*)$, $p(V - 2(\underline{\beta} - \gamma)) = pS*+\bar{c}$

G principal = $pS*+\bar{c} - c + (1-p)S* = S*+\bar{c}*-c$  (4)

## 3 CONCLUSION

In this model, the principal is planning building electronic commerce with some of her business partners. Thus, the contracted supplier's task consists in the adoption of some ICT business tool. For instance, the firm can incorporate to its website the ability to buy by mail order or additionally to accept online payments through an automated process.

The results of the paper follow directly from [2]. Once network effects are introduced to the model, agent's desutility reduces leading to an increase in social surplus.

## APPENDICES

Proof proposition II:

$p(T - 2(\underline{\beta} - \gamma)) + (1-p)(T - 2(\bar{\beta} - \gamma))$

$= T - 2p(\underline{\beta} - \gamma - (\bar{\beta} - \gamma)) - 2(\bar{\beta} - \gamma)$

$= T - 2(p(\underline{\beta} - \gamma) - p(\bar{\beta} - \gamma) + (\bar{\beta} - \gamma))$

$= T - 2(p(\underline{\beta} - \gamma) + (1-p)(\bar{\beta} - \gamma)) = T - 2(E\beta - \gamma)$

Proof Footnote 1:

$T = 2(E\beta - \gamma)$

Low cost agent's benefit = $T - 2\underline{\beta} = 2(p\underline{\beta} + (1-p)\bar{\beta}) - 2\gamma - 2\underline{\beta}$

$T - 2\underline{\beta} = 2(1-p)(\bar{\beta} - \underline{\beta}) - 2\gamma$

From proposition II's contract, c = p(T-Desutility). Then

c' = p(T - 2$\underline{\beta}$), c' = $2p(1-p)(\bar{\beta} - \underline{\beta}) - 2\gamma p$.

Proof proposition III:

$[p+(1-p)p](V-2\underline{\beta})+(1-p)^2(v-\bar{\beta})+\gamma[p+(1-p)p+1] =$

$[p+(1-p)p](V-2\underline{\beta})+(1-p)^2(v-\bar{\beta})+\gamma[p+(1-p)p+1] =$

$p(V-2\underline{\beta})+(1-p)p(V-2\underline{\beta})+(1-[p+(1-p)p])(v-\bar{\beta})+\gamma(1+p)+\gamma(1-p)p =$

$p(V-2\underline{\beta})+(1-p)p(V-2\underline{\beta})+(v-\bar{\beta})(1-p)-(1-p)p(v-\bar{\beta})+\gamma(1+p)+\gamma(1-p)p =$

$p(V-2\underline{\beta})+(1-p)(v-\bar{\beta})+(1-p)p(V-2\underline{\beta}-(v-\bar{\beta}))+\gamma(1+p)+\gamma p(1-p)$

$= S*+\bar{c}*$

**M. V. Alderete** Ph. D in Economics, expected. CONICET (National Commission on Scientific and Technological Research in Argentina)'s Postgraduate Scholarship holder. Teaching assistant at the Department of Economics, Universidad Nacional Del Sur. Bahía Blanca, Argentina.
Empirical research grant from Fundación Observatorio PyME 2005 in Industrial Economy: "Building networks as a competitive advantage in small and medium sized enterprises (SMEs)."